\RequirePackage{snapshot}            
\documentclass[letterpaper,twoside]{article}
\usepackage{latexsym,times}
\usepackage[intlimits]{amsmath}
\usepackage{amsfonts,amssymb}
\DeclareSymbolFontAlphabet{\mathbb}{AMSb}
\usepackage[mathscr]{eucal}
\usepackage[notoday]{rcsinfo}
\usepackage{changebar} 
\usepackage[textstyle,cdot,squaren,thickspace,thickqspace]{SIunits}
\usepackage[colon,authoryear]{natbib}
\usepackage{fancyvrb}
\usepackage{url}
\usepackage{xspace}
\usepackage[usenames,dvips]{color}
\usepackage[dvips]{graphicx}
\DeclareGraphicsExtensions{.eps}   
\graphicspath{
  {./}
}
\usepackage[letterpaper,top=1in,bottom=1in,hmargin={1in,1in}]{geometry}
\numberwithin{equation}{section}
\usepackage{float}
\usepackage[bf,float,subfigure]{caption2}
\setcaptionmargin{0.5in}
\usepackage[sf]{subfigure}

\makeatletter
\DeclareRobustCommand\em
{\@nomath\em \ifdim \fontdimen\@ne\font >\z@
  \upshape \else \slshape \fi}

\makeatother

\begin{document}
\title{Perceptions of Complex Systems Are Governed by Power Laws}
\author{Uygar \"{O}zesmi}
\date{}
\maketitle

\noindent
Department of Environmental Engineering, Erciyes University, 38039 Kayseri, Turkey, Uygar@ozesmi.org
\rcsInfo $Id: template.tex,v 1.3 2004/06/30 23:26:18 jonnyreb Exp $
\typeout{}
\typeout{<==\rcsInfoFile\ \rcsInfoRevision\ \rcsInfoDate\ \rcsInfoOwner==>}
\typeout{}
\begin{abstract}
Many networks in natural and human-made systems exhibit scale-free properties and are small worlds. Now we show that people's understanding of complex systems in their cognitive maps also follow a scale-free topology. People focus on a few attributes, relating these with many other things in the system. Many more attributes have very few connections. People use relatively short explanations to describe events; their cognitive map is a small world with less than six degrees of separation. These findings may help us to better understand people's perceptions, especially when it comes to decision-making, conflict resolution, politics and management.
\end{abstract}


\noindent
\Large I\normalsize{n} nature many networks from protein interaction \cite{1} to metabolism \cite{2} show scale-free properties.  In human systems scale-free properties have been observed in the world-wide web \cite{3}, the internet \cite{4}, linguistics \cite{5}, sexual contacts \cite{6}, movie actor collaboration \cite{7}, and scientific collaboration \cite{8}. Here we show that people's perceptions of ecosystems and other complex systems also obey power laws.

We examined people's perceptions of complex systems with the technique of cognitive mapping. Cognitive maps are networks that have weighted and directed edges (causal connections of varying strengths) between nodes (variables). Axelrod \cite{9} first used signed digraphs to represent causal relationships among variables as defined and described by people and he called these representations cognitive maps. By using weighted connections instead of binary ones, Kosko \cite{10} defined fuzzy cognitive maps. Cognitive maps have been used to examine decision-making \cite{11}, people's perceptions of complex social systems \cite{12}, and for modelling in various fields including operations management \cite{13}, virtual reality \cite{14} and environmental management \cite{15}. Eden et al. \cite{16} have extensively used cognitive mapping to examine decision-making and problem-solving in businesses. Recently, fuzzy cognitive maps created with expert knowledge have been used in data mining of the world wide web \cite{17}. 

\section*{Methods}

\noindent
\textbf{Obtaining cognitive maps.} We have done in-depth interviews with people about complex systems. First the process of drawing a fuzzy cognitive map was shown with a completely unrelated map.  Then the stakeholders (individuals, couples, or small groups of up to 5 people) were asked to draw their own cognitive map in response to open-ended questions such as "What variables come to mind if I mention -insert a system-, how do these variables affect each other?". The stakeholders listed the important variables.  They signified the relationships between these variables by drawing lines between them and using arrows to indicate the directions of the relationships. They also gave them signs of positive or negative, and strengths of a lot (1), some (0.5), or a little (0.25). After the stakeholders drew their cognitive maps, which are essentially directed weighted graphs, they were coded into adjacency matrices. These cognitive maps were then augmented and added together \cite{18} to create a social cognitive map of stakeholder groups or of all the stakeholders interviewed for each study system. The interviewing method and analyses are described in detail by \"{O}zesmi and \"{O}zesmi \cite{19}.

\noindent
\textbf{Structure of cognitive maps.} We analyzed the structure of these cognitive maps by examining the outdegree, the indegree, and total degree (centrality) of the variables.  Outdegree shows the total strength of the connections exiting from a variable:

\begin{equation*}
\mathrm{od}(v_i) = \sum_{k-1}^N \bar{a}_{ik}
\end{equation*}

\noindent
Indegree shows the total strength of the connections coming into a variable:

\begin{equation*}
\mathrm{id}(v_i) = \sum_{k-1}^N \bar{a}_{ki}
\end{equation*}

\noindent
Total degree (centrality) is the sum of indegree and outdegree of a variable:

\begin{equation*}
c_i = \mathrm{td}(v_i) = \mathrm{od}(v_i) + \mathrm{id}(v_i) 
\end{equation*}

Total degree shows the cumulative strength of connections entering into and exiting from a variable.  It indicates how important a variable is in the map.

We examined the average distance between variables assuming that the connections are undirected.  

We also looked at other structural indices to determine if maps from different study systems were similar.  We calculated the ratio of receiver to transmitter variables ($R/T$).  Receiver variables have a positive indegree, $\mathrm{id}(v_i)$, and zero outdegree, $\mathrm{od}(v_i)$. Transmitter variables have a positive outdegree, $\mathrm{od}(v_i)$, and zero indegree, $\mathrm{id}(v_i)$. More complex maps will have larger ratios of receiver to transmitter variables because they define more utility outcomes and less controlling forcing functions.

We examined the density (clustering coefficient) of a fuzzy cognitive map ($D$), calculated as the number of connections divided by the maximum number of connections possible between $N$ variables \cite{20}: 

\begin{equation*}
D = \frac{k}{N^2}
\end{equation*}

Density is an index of connectivity that shows how connected or sparse the maps are. If the density of a map is high then then there are a large number of causal relationships among the variables.

Another structural measure of a cognitive map is the hierarchy index ($h$) \cite{21}:

\begin{equation*}
h = \frac{12}{(N-1)N(N+1)} \sum_i \left[ \mathrm{od}(v_i) - \frac{\sum \mathrm{od}(v_i)}{N} \right]^2
\end{equation*}
 
\noindent
where $N$ is the total number of variables.  When $h$ is equal to zero the system is fully democratic and when $h$ is equal to one then the map is fully hierarchical.

\section*{Results}

\noindent
\textbf{Structure of cognitive maps.} We found out that if a standard methodology is used the structural indices of the social cognitive maps from separate study systems are in the same range (Table \ref{tab1}).  Typically there are about 4 connections per variable in the maps.  The number of transmitter variables was almost always higher than the number of receiver variables in the maps.  This indicates that people perceived the systems as having more forcing functions than utility variables. The density of the maps are low, indicating maps with relatively few relationships among variables.  The hierarchy index of the maps indicates that the maps tend to be more "democratic" than "hierarchical" in their structure.

\noindent
\textbf{Total degree.} When we examined the total degree of the variables in individual stakeholder cognitive maps, they had power, Poisson, bimodal or uniform distributions. However, in social cognitive maps, the degree distribution of the total degree, or centrality, always follows a power law. The total degree exponents, $\gamma_{\mathrm{total}}$, vary from -0.995 to -1.546 for our six studies with $R^2$ between 0.94 to 0.98 (Table \ref{tab2}). Because cognitive maps have directed links (i.e. agricultural runoff causes lake eutrophication), we examined the histograms of the indegree and outdegree. These also followed a power law distribution for the social cognitive maps (Table \ref{tab2}). The indegree exponents, $\gamma{_\mathrm{in}}$, vary from -1.004 to 1.530 ($R^2$ = 0.93 - 0.99). The outdegree exponents, $\gamma_{\mathrm{out}}$, vary from -1.292 to -1.890 ($R^2$ = 0.87 - 0.97).

\noindent
\textbf{Average distance.} As in most real world networks \cite{22} social cognitive maps are small worlds, with an average distance ($l$) between nodes varying from 3 to 5 (Table \ref{tab2}). These distances are similar to the distances in random networks of the same size. Although the causal links between variables are directed, the average distances were calculated assuming that the links between nodes were bidirectional. Therefore we would expect the lengths of causal chains of reasoning to be even shorter on average. In other words, people's perceptions of complex systems involve causal chains that are on average less than 5 links long. One could assume that the longer the causal chain the greater level of detail that people use to explain an event.

\section*{Discussion}

\noindent
\textbf{Scale-free networks.} Many large networks have the property that the vertex connectivities follow a scale-free power-law distribution because: 1) networks expand continuously by the addition of new vertices, and 2) new vertices attach preferentially to sites that are already well-connected \cite{7}. With social cognitive maps, as more individual maps are added to the social map, the map expands with new nodes and new connections. In addition, as people in the same geographic area or social group tend to have similar perceptions about systems or problems, they think of the same variables and causal relationships between those variables. Therefore the more maps that are added together, the stronger the causal connections become between shared variables.

Social cognitive maps were found to be small worlds with variable connectivities following a scale-free power-law distribution. These results have implications for how people perceive systems and could help solve problems in many different areas such as cognition, perception, decision-making, conflict resolution, politics, and management. For example, by analyzing networks of social interactions together with cognitive maps we may be better able to understand how ideas spread and become accepted by group members. Finally cognitive maps of people's perceptions of complex systems provide another example of networks with scale-free topology, presenting a case where the network connections are directed and weighted.

\small
\noindent
This research was in part funded by the Turkish Scientific and Technical Research Council (TUBITAK-YDABAG), the Turkish Society for the Protection of Nature (DHKD), Turkish State Hydraulic Works, and the University of Minnesota MacArthur Program.

\normalsize

\bibliography{powerLaw}
\bibliographystyle{unsrt}

\newpage

\begin{table}
\begin{center}
\caption{Values for the number of individual maps in the social cognitive map ($n$), number of variables ($N$), number of edges or causal connections ($k$), and graph theory indices, receiver to transmitter ratio ($R/T$), density ($D$), hierarchy ($h$), for social cognitive maps from six different study systems.}
\label{tab1}
\begin{tabular}{p{1in}lllllll}
System&$n$&$N$&$k$&$k/N$&$R/T$&$D$&$h$\\
\hline
Kizilirmak Delta \par ecosystem&31&136&616&4.616&0.333&0.033&0.026\\
Yusufeli dam \par construction&14&97&360&3.711&0.083&0.038&0.049\\
Uluabat Lake \par ecosystem&35&253&1173&4.636&0.116&0.018&0.011\\
Kayseri \par industry&30&135&948&7.02&0.64&0.050&0.080\\
Sultan Marshes \par ecosystem&56&181&773&4.27&1.03&0.024&0.118\\
Tuzla Lake \par ecosystem&44&204&864&4.24&0.39&0.021&0.024\\
\end{tabular}
\end{center}
\end{table}

\newpage
\begin{table}
\begin{center}
\caption{Total degree, indegree and outdegree exponents with the $R^2$ values for power law distribution together with the average distances, l, of social cognitive maps from six different study systems.}
\label{tab2}
\begin{tabular}{p{1in}llllllllllll}
System&$n$&$N$&$k$&$\gamma_\mathrm{total}$&$R^2$&$\gamma_\mathrm{in}$&$R^2$&$\gamma_\mathrm{out}$&$R^2$&$l_\mathrm{rand}$&$l$&Ref.\\
\hline
Kizilirmak Delta \par ecosystem&31&136&616&-1.127&0.98&-1.084&0.95&-1.373&0.87&3.25&3.72&\cite{18}-\cite{19}\\
Yusufeli dam \par construction&14&97&360&-1.238&0.97&-1.004&0.94&-1.345&0.94&3.49&3.78&\cite{20}\\
Uluabat Lake \par ecosystem&35&253&1173&-1.546&0.98&-1.530&0.99&-1.890&0.97&3.61&3.77&\cite{21}\\
Kayseri \par industry&30&135&948&-0.995&0.95&-1.257&0.93&-1.537&0.94&2.52&3.06&\cite{22}\\
Sultan Marshes \par ecosystem&56&181&773&-1.213&0.94&-1.312&0.99&-1.292&0.97&3.58&4.69&\cite{23}\\
Tuzla Lake \par ecosystem&44&204&864&-1.500&0.97&-1.386&0.96&-1.857&0.94&3.68&4.92&\cite{24}\\
\end{tabular}
\end{center}
$N$ is the number of variables in the social cognitive map, $k$ is the number of edges (causal connections), and $n$ is the number of maps added together to create the social cognitive map.
\end{table}

\newpage

\begin{figure}
\caption{Cumulative distributions of total degree of social cognitive maps from six different complex systems.}
\label{fig1}
\includegraphics{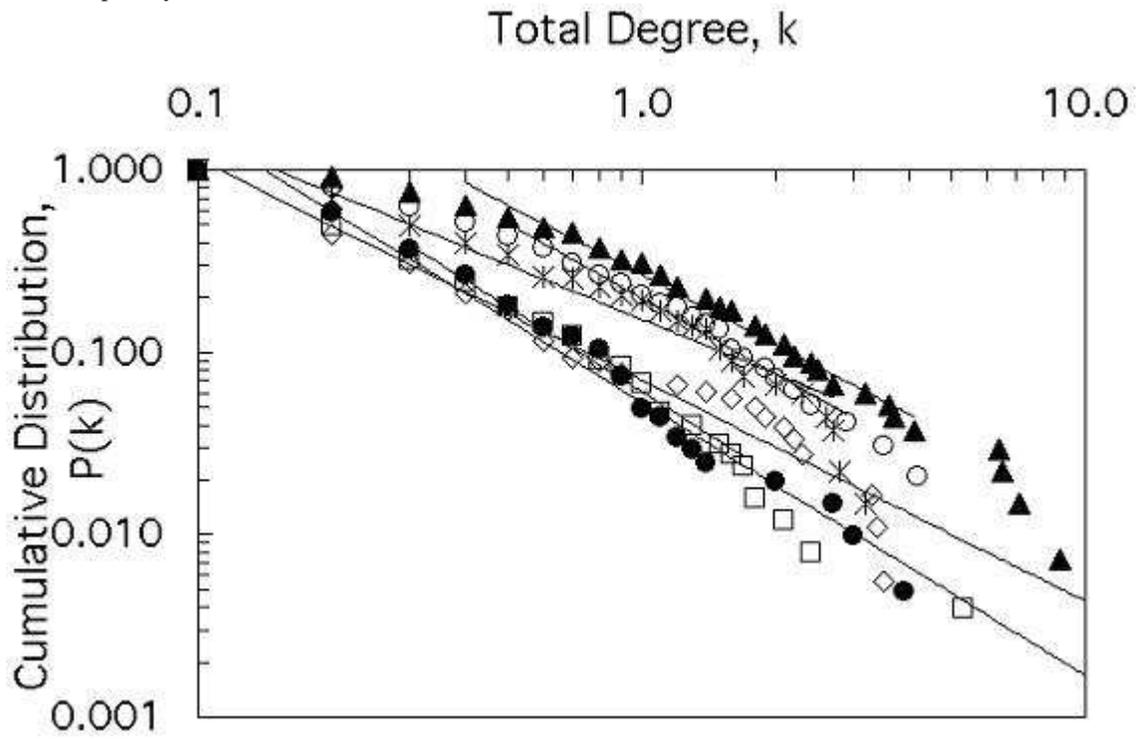}
\end{figure}

\end{document}